# *Size effects and idealized dislocation microstructure at small scales: predictions of a phenomenological model of Mesoscopic Field Dislocation Mechanics: Part I*

Amit Acharya[*] and Anish Roy

Civil and Environmental Engineering, Carnegie Mellon University, Pittsburgh, PA 15213, U.S.A

**Abstract**

A Phenomenological Mesoscopic Field Dislocation Mechanics (PMFDM) model is developed, extending continuum plasticity theory for studying initial-boundary value problems of small-scale plasticity. PMFDM results from an elementary space-time averaging of the equations of Field Dislocation Mechanics (FDM), followed by a closure assumption from any strain-gradient plasticity model that attempts to model effects of geometrically-necessary dislocations (GND) only in work-hardening. The specific lower-order gradient plasticity model chosen to substantiate this work requires one additional material parameter compared to its conventional continuum plasticity counterpart. *The further addition of dislocation mechanics requires no additional material parameters*. The model a) retains the constitutive dependence of the free-energy only on elastic strain as in conventional continuum plasticity with no explicit dependence on dislocation density, b) does not require higher-order stresses, and c) does not require a constitutive specification of a 'back-stress' in the expression for average dislocation velocity/plastic strain rate. However, long-range stress effects of average dislocation distributions are predicted by the model in a mechanistically rigorous sense. Plausible boundary conditions (with obvious implication for corresponding interface conditions) are discussed in some detail from a physical point of view. Energetic and dissipative aspects of the model are also discussed. The developed framework is a continuous-time model of averaged dislocation plasticity, without having to rely on the notion of incremental work functions, their convexity properties, or their minimization. The tangent modulus relating stress rate and total strain rate in the model is the positive-definite tensor of linear elasticity, and this is not an impediment to the development of idealized microstructure in the theory and computations, even when such a convexity property is preserved in a computational scheme. A model of finite deformation, mesoscopic single crystal plasticity is also presented, motivated by the above considerations.

Lower-order gradient plasticity appears as a constitutive limit of PMFDM, and the development suggests a plausible boundary condition on the plastic strain rate for this limit that is appropriate for the modeling of constrained plastic flow in three-dimensional situations.

## 1. Introduction

In our view, the main challenges of a model of plasticity at the micron scale, building upon conventional, size-independent, macroscopic plasticity, are the accurate representation of

1. the stress field of average, signed dislocation density (GND) that does not vanish at the scale of resolution of such models,
2. the plastic strain rate arising from the temporal evolution of this density, and

---

[*] Corresponding author: Tel. (412) 268 4566; Fax. (412) 268 7813; email: acharyaamit@cmu.edu

3. the effect of GNDs on the strength of the material.

These features of plastic response at this scale lead to size effects and the development of microstructure under macroscopically homogeneous conditions; any mechanical model ought to strive to account for these features in as mechanistically rigorous a manner as possible. The scales of spatial and temporal resolution of such models, however, are large enough that accounting for individual dislocations is not possible. A minor and obvious complicating factor is that the spatial averaging cancels pairs of positive and negative dislocations but the plastic strain rate produced from the motion of such pairs (as in loop expansion) does not cancel and needs to be accounted for in the model. Major, and subtle, complicating factors are the nonlinearity of the underlying dislocation dynamics and the large reduction in degrees of freedom implied by the desired coarse-grained model, resulting in memory-dependent response and stochastic behavior of averaged variables of an autonomous, deterministic fine-scale theory, as can be shown for low-dimensional dynamical systems (Acharya, 2005; Sawant and Acharya, 2004).

With the above thoughts in mind, we find it best to start from the averaging of an appropriate model of dislocation mechanics, seeking to define a closed model for the averaged variables. Since the operations of averaging and nonlinear combinations of variables do not commute, we are inevitably faced with a model of averaged variables that requires input from the finer scale model. In this work, we take recourse to phenomenologically specifying these inputs, the main input involved being a model of plastic straining due to the dislocations that are averaged out (statistically-stored dislocations or SSDs). For this purpose, we borrow a robust macroscopic plasticity model, suitably coupled to the averaged GND mechanics. Despite the phenomenology, the averaging procedure and the fine-scale theory involved impart to the coarse model a rich structure that enables a gamut of relevant predictions, with only one extra material parameter over and above conventional macroscopic continuum plasticity. This paper and its sequel provide a detailed account of this development and its predictions. A feature of our model that may be considered a desirable consistency property is that when the spatial averaging length scale tends to zero, our model reduces to a theory of the collective behavior of individual dislocations – applicable to finite bodies of arbitrary shape, and conveniently adapted for geometric, crystal elastic and dissipative nonlinearities as well as the effects of inertia - that can be exercised to probe mechanical behavior at the nanoscale and above, with emerging predictions representative of such (Acharya, 2001, 2003, 2004; Miller and Acharya, 2004; Roy and Acharya, 2005; Varadhan et al., 2005).

Mechanical models of size-dependent plasticity and plastic heterogeneity at the micron scale for solving initial-boundary value problems have been the subject of much current research (Aifantis, 1984; Fleck and Hutchinson, 2001; Acharya and Bassani, 1996, 2000; Acharya et al., 2004; Busso et al. 2000; Menzel and Steinmann, 2000; Gao et al., 2000; Shu et al., 1999, 2001; Gurtin, 2000; Bittencourt et al., 2003; Arsenlis et al., 2002; Evers et al., 2004; Ortiz and Repetto, 1999, Shizawa and Zbib, 1999, Svendsen, 2002, Gudmundson, 2004; Aifantis and Willis, 2004; Gurtin and Anand, 2005). The concepts and methodological details of our effort are different from the works mentioned above for reasons mentioned in the abstract of this paper.

Somewhat similar approaches to the present work are those of Arsenlis et al. (2004), El-Azab (2000, 2005), and Yefimov et al. (2004). While the primary field equations of El-Azab and Yefimov et al. are motivated by procedures in statistical mechanics, the standard problem of closure in statistical mechanical treatments for nonlinear theories – related to the *derivation* of correlations in space and time resulting in memory and spatially non-local effects - is not resolved in these treatments but modeled, thus rendering them phenomenological. All of these approaches prefer working with slip system dislocation density variables, as was also suggested in Acharya (2003). Physically reasonable models for the slip system dislocation density fluxes are developed, both for GNDs and SSDs; noteworthy amongst these is a PDE for SSD evolution inferred in the works of Groma (1997) and Groma and Balogh (1999). However, it must be realized that a physical conservation statement for Burgers vector content leads to an evolution law for only the total dislocation density tensor (Nye tensor); any division of this conservation statement into numerous balances for slip system dislocation densities, even though based on physical premises, is necessarily ad-hoc (e.g. El-Azab, 2000, 2005). Additionally, the definition of slip system densities also results in operational ambiguities related to the definition of initial conditions. For example, a screw dislocation density in an FCC crystal cannot be unambiguously attributed to any particular slip system whose slip direction is collinear with the Burgers vector of the screw density. Given the extra field variables that are introduced in going to a slip-system density based formulation for dislocation density evolution as compared to Nye tensor evolution, a practical option may very well be to evolve only the Nye tensor but use suitable components of it during the course of evolution to define slip-system GND fluxes. One such option is developed in this paper. Also from the practical point of view, solving partial differential equations for phase space densities (while accounting for fluctuations) involves an increase in the number of *independent* variables, and this is known to be computationally expensive; moreover, at many points of space and time the solution of specific problems may be expected to be deterministic in which case the true solution for the phase density is a Dirac-delta function in appropriate phase variables. Resolving such functions in actual computations is not a straightforward matter.

With respect to the calculation of the initial and evolving stress field in these approaches, the standard approach of continuum plasticity related to defining a plastic strain rate along with an appropriate decomposition of the total strain rate, the equilibrium equations, and an elastic constitutive assumption is used. When stresses of dislocation density distributions have to be recovered, it is our belief – and one that is substantiated in a later section of this paper and through a computational result in Part II of this paper – that there is much to be gained by retaining the structure of equations of the elastic theory of continuously distributed dislocations (ECDD) (Kroner, 1981; Willis, 1967). This feature is not preserved by the technique mentioned in the first line of this paragraph, borrowed from conventional plasticity. Moreover, accommodating ECDD within a nonequilibrium theory of plasticity takes some care (Acharya, 2003), especially in the finite deformation setting (Acharya, 2004).

For the important reason that there is no physical criterion to choose a fixed, coherent reference configuration for plasticity in the finite deformation setting, any statement of evolution of dislocation density ought to be developed in terms of appropriate density measures and spatial derivatives on the current configuration. Otherwise, the physical invariance, with respect to choice of reference configuration, of any such statement ought to be demonstrated. Such invariance is not demonstrated in the work of Arsenlis et al. (2004); an argument that in effect could be construed as illustrating this invariance is attempted in El-Azab (2005), but we are unable to judge its correctness from the mathematical details that are provided. Additionally, if an additive decomposition of the velocity gradient (or a multiplicative decomposition of the deformation gradient) is assumed along with a constitutive equation for the plastic part of the velocity gradient, then the usual procedures of continuum plasticity demonstrate that the choice of a deformation history completely determines the evolution of the field $\boldsymbol{F}^e$ (elastic distortion tensor in the multiplicative decomposition). For the prediction of the internal stress field of the dislocation distribution correctly, it is essential that this $\boldsymbol{F}^e$ field be consistent with the incompatibility equation linking appropriate spatial derivatives of some function of $\boldsymbol{F}^e$ to the Nye tensor. Depending on the details of each model, the latter requirement may be used as a constraint in defining the dislocation density evolution (Arsenlis et al., 2004), or the conventional constitutive prescription of the plastic part of the velocity gradient has to be given up (Acharya, 2004). This important issue is left unresolved in El-Azab (2005).

A somewhat different approach to the study of plasticity at these scales, with impressive predictions, is the work of Koslowski et al. (2002), applied in Koslowski et al. (2004). The solution of problems involving boundary conditions on finite bodies would not seem – or, has not yet been demonstrated - to be within the scope of this model with its usual efficiency. The primary conceptual difference between the work of Koslowski et al. (2002, 2004) and our work is that the plastic distortion in their model evolves discretely in time by minimization of the incremental work of deformation. Apparently, no information regarding the tensorial kinematics of dislocation *motion* is important for macroscopic behavior and microstructure development other than prescribing a simple dissipative work contribution to the incremental work functional. Minimization of incremental work appears to be the controlling physical principle, notwithstanding the issue of non-uniqueness of (relative) minimizers. In contrast, in our work the rate of plastic distortion arises from an averaging of a fine scale theory (FDM) that incorporates the elastic theory of dislocations exactly; as for kinetics, some of the simplest realizations of the equation for dislocation density-evolution in FDM have been shown (Acharya 2003; Varadhan et al. 2005) to lead to Hamilton-Jacobi equations for propagation of fronts, whose solutions can be interpreted as expansions of loops from Frank-Read sources allowing automatic annihilation. Moreover, features like non-Schmid effects in dislocation mobility and initiation of cross-slip find a mechanistic status within such a kinetic framework (Acharya, 2003)[i], derived from the tensorial kinematics of dislocation motion. More importantly, in this paper we find that averaging such equations leaves a significant trace of the fine-scale kinematics in the averaged equations, which is also physically reasonable considering the length-scales involved. What connections, if any, might exist between minimization of incremental work-driven kinetics and kinetics derived from Field Dislocation Mechanics remains to be understood by us.

Some interesting ideas related to coarse-graining procedures that may prove to be of utility in modeling particular aspects of mesoscopic plasticity are presented in LeSar and Rickman (2003). The fundamental challenge of coarse-graining a nonlinear *dynamical* system like dislocation dynamics, however, remains unaddressed along with the delineation of the circumstances in which coarse-grained energetics can be of significance to the understanding of coarse-grained dynamics. A (rigorous) example, related to materials science, of the failure of dynamics derived from averaged energetics to provide meaningful information for averaged dynamics can be found in Abeyaratne et al. (1996). There are strong reasons to believe that mesoscopic plasticity will be no exception to the 'principle' demonstrated in Abeyaratne et al. (1996) in general, and so a clear understanding of the nature of the approximation involved in adopting averaged-energy driven gradient flow dynamics in the case of collective dislocation behavior could be quite useful, both practically and conceptually.

---

[i] While the emergence of non-schmid effects were demonstrated in the context of individual dislocation behavior, essentially the same arguments hold for macroscopic (averaged) non-schmid effects with the added assumption that microscopic dislocation motion initiates when a scalar valued-function of the microscopic driving force achieves a material specific threshold (microscopic yield). Any physical, scalar-valued function of the driving force can only depend on its magnitude. This eliminates any dependence on sign of microscopic Burgers vector in the macroscopic spatial average of the microscopic yield criterion, while retaining the dependence of the microscopic yield criterion on non-schmid stress components arising due to special core displacement (Burgers vector) geometries.

## 2. Averaged Field Dislocation Mechanics, Boundary Conditions, and Initial Conditions

The field equations of FDM may be written as follows[ii]:

$$\begin{aligned}
&curl\ \chi = \alpha \\
&div\ \chi = 0 \\
&div(grad\ \dot{z}) = div(\alpha \times V) \\
&div\left[C:\{grad(u-z)+\chi\}\right] = 0 \\
&\dot{\alpha} = -curl(\alpha \times V) + s.
\end{aligned} \quad (1)$$

Here, $\chi$ is the incompatible part of the elastic distortion tensor, $u$ is the total displacement field, $u-z$ is a vector field whose gradient is the compatible part of the elastic distortion tensor, $C$ is the fourth-order, possibly anisotropic, tensor of linear elastic moduli, $\alpha$ is the dislocation density tensor, $V$ is the dislocation velocity vector, and $s$ is a dislocation nucleation rate tensor (not related to dislocation line length increase from *existing* dislocations). The argument of the $div$ operator in $(1)_4$ is the stress tensor, and the functions $V$ and $s$ are constitutively specified. These constitutive specifications are generally nonlinear in the basic fields. The physical motivation behind (1) is to be found in Acharya (2001, 2004). Equation $(1)_3$ is the differential equation corresponding to the variational statement defining $\dot{z}$ as the potential field whose gradient is the compatible part of $\alpha \times V$ (Stokes-Helmholtz resolution). The actual solution of the above system requires, of course, boundary and initial conditions.

In the following, we adapt a commonly used averaging procedure utilized in the study of multiphase flows (e.g. Babic, 1997) for our purposes. For a microscopic field $f$ given as a function of space and time, we define the mesoscopic space-time averaged field $\bar{f}$ as follows:

$$\bar{f}(x,t) := \frac{1}{\int_{I(t)}\int_{\Omega(x)} w(x-x',t-t')\ dx'dt'} \int_{\Im}\int_{B} w(x-x',t-t') f(x',t')\ dx'dt', \quad (2)$$

where $B$ is the body and $\Im$ a sufficiently large interval of time. In the above, $\Omega(x)$ is a bounded region within the body around the point $x$ with linear dimension of the order of the spatial resolution of the macroscopic model we seek, and $I(t)$ is a bounded interval in $\Im$ containing $t$. The averaged field $\bar{f}$ is simply a weighted, space-time, running average of the microscopic field $f$. The weighting function $w$ is non-dimensional, assumed to be smooth in

---

[ii] The cross product of a tensor $A$ and a vector $b$ is defined as $(A \times b)c = (A^T c) \times b\ \forall$ vectors $c$. The *curl* of a tensor field is defined as $(curl\ A)c = curl(A^T c)\ \forall$ spatially uniform vector fields $c$.

the variables $x, x', t, t'$ and, for fixed $x$ and $t$, have support (i.e. to be non-zero) only in $\Omega(x) \times I(t)$ when viewed as a function of $(x', t')$.

### *2.1 Averaged Field Equations*

We now apply the averaging operator (2) to each side of all the equations of (1), thinking of the space and time variables involved in (1) as $x', t'$. *The elastic moduli are assumed to be spatially uniform* for simplicity, as the interesting features of macroscopic plastic response do not arise from elastic heterogeneity. Using the relations

$$\frac{\partial w}{\partial x'_i} = -\frac{\partial w}{\partial x_i}, \qquad \frac{\partial w}{\partial t'} = -\frac{\partial w}{\partial t} \qquad (3)$$

it is not difficult to see that for almost all $(x,t) \in B \times \Im$ (except for boundary layers in both variables), $(1)_{1,2,4}$ retain the same form, but now stated in terms of the fields $\bar{u}, \bar{\chi}, \bar{z}$ and derivative operators in terms of the variables $x, t$. Defining

$$L^p(x,t) := \overline{(\alpha - \bar{\alpha}) \times V}(x,t) = \overline{\alpha \times V}(x,t) - \bar{\alpha}(x,t) \times \bar{V}(x,t), \qquad (4)$$

equations $(1)_{3,5}$ may be written in the form

$$div(grad\,\dot{\bar{z}}) = div(\bar{\alpha} \times \bar{V} + L^p) \qquad (5)$$

and

$$\dot{\bar{\alpha}} = -curl(\bar{\alpha} \times \bar{V} + L^p) + \bar{s}. \qquad (6)$$

Henceforth, we refer to $\bar{\alpha} \times \bar{V} + L^p$ as the macroscopic *slipping distortion* $S$.[iii]

Clearly, the averaged equations do not form a closed set because of the terms $\bar{V}, \bar{s}, L^p$ whose representation in terms of the averaged fields, if such were to exist, is not known. The central, non-trivial, coarse-graining problem in this context is the determination of the time evolution of the fields $\bar{V}, \bar{s}, L^p$ in terms of the fields $\bar{\chi}, \bar{z}, \bar{u}, \bar{\alpha}$, most likely involving dependencies on the memory of the latter and stochastic effects (Sawant and Acharya, 2004).

The abovementioned coarse-graining question is not the one we address in this paper. Instead we simply phenomenologically model the required terms. Of course, the predictions we make with the so-developed model depend upon the *structure* of the averaged equations and associated boundary conditions in an essential way, and go much beyond what can be inferred solely from knowledge of the introduced phenomenology.

---

[iii] There is a reason for not referring to this term as the plastic part of the (small deformation) velocity gradient that will be discussed in the next section.

In order to aid the specification of constitutive assumptions, we examine the plastic working within the model. The averaged *elastic distortion* is given by

$$\bar{U}^e = grad(\bar{u} - \bar{z}) + \bar{\chi} \tag{7}$$

and the average *plastic distortion* as

$$\bar{U}^p := grad\,\bar{u} - \bar{U}^e = grad\,\bar{z} - \bar{\chi}. \tag{8}$$

The rate of plastic working is

$$\int_B \bar{T} : \dot{\bar{U}^p}\, dv, \tag{9}$$

and we would like to uncover the entities power-conjugate to the terms $\bar{V}$, $L^p$, and $\bar{s}$. To do so, we assume the average stress field to be smooth and invoke an orthogonal decomposition of the field akin to the Stokes-Helmholtz resolution satisfying special boundary conditions (Weyl, 1940; Acharya 2004):

$$\bar{T} = curl\, W_{\bar{T}} + grad\, g, \tag{10}$$

where $W_{\bar{T}}$ vanishes on the boundary of the body. Now, using (8) in the form of rates in (9) and invoking (10), the averaged equations corresponding to $(1)_{1,2}$, and (5), (6), we find that

$$\int_B \bar{T} : \dot{\bar{U}^p}\, dv = \int_B \bar{T} : L^p\, dv + \int_B \xi \cdot \bar{V}\, dv + \int_B (-W_{\bar{T}}) : \bar{s}\, dv$$
$$\xi := X(\bar{T}\bar{\alpha}) \quad;\quad \xi_i = e_{ijk}\bar{T}_{jr}\bar{\alpha}_{rk}, \tag{11}$$

where $X$ is the third-order alternating tensor. In deriving (11), the boundary conditions

$$\dot{\bar{\chi}} n = 0$$
$$\left( grad\, \dot{\bar{z}} - \bar{\alpha} \times \bar{V} - L^p \right) n = 0 \tag{12}$$

on the entire boundary $\partial B$ of the body have been used (these are boundary conditions that are used in defining our coarse model, as can be seen subsequently from $(23)_{3,5}$). Thus, $\bar{T}, \xi$, and $-W_{\bar{T}}$ are designated as the driving forces for the inelastic mechanisms $L^p$, $\bar{V}$, and $\bar{s}$.

The variable $\bar{V}$ has the obvious physical meaning of being a space-time average of the pointwise, microscopic dislocation velocity. By definition, the tensor field $L^p$ can be seen to be representative of a portion of the average slip strain rate produced by the 'microscopic' dislocation density; in particular, it can be non-vanishing even when $\bar{\alpha} = 0$ and, as such, it is to be physically interpreted as the strain-rate produced by so-called 'statistically-stored dislocations' (SSD), as is also indicated by the extreme right-hand side of (4). An elementary, idealized, realization of such a situation corresponds to a uniformly expanding square loop within $\Omega(\boldsymbol{x})$. Here $\boldsymbol{b}$ is assumed to be the Burgers vector density per unit area, uniform along the loop, and $\boldsymbol{l}$ the unit line direction at each point of the loop; opposite edges of the loop cancel each other in the spatial averaging yielding $\bar{\boldsymbol{\alpha}} = \boldsymbol{0}$. The velocity is assumed to lie in a slip plane and point outwards with respect to the loop with uniform magnitude; along individual parallel sides of the loop with opposite direction

$$\boldsymbol{\alpha} \times \boldsymbol{V} = \boldsymbol{b} \otimes (\boldsymbol{l} \times \boldsymbol{V}) \tag{13}$$

is identical (since $\boldsymbol{l}$ and $\boldsymbol{V}$ both change sign going from one side to the other) and hence

$$\boldsymbol{L}^p = \overline{\boldsymbol{\alpha} \times \boldsymbol{V}} - \bar{\boldsymbol{\alpha}} \times \bar{\boldsymbol{V}} \neq \boldsymbol{0}. \tag{14}$$

Finally, $\bar{\boldsymbol{s}}$ is the averaged nucleation rate tensor, expected to be small in most cases due to spatial averaging over the mesoscopic volume $\Omega(\boldsymbol{x})$.

*2.2 Boundary conditions for averaged field equations*

Boundary conditions for a mathematical model may be based on definite knowledge of the details of such conditions from the physical phenomena being modeled, when such knowledge is available. Often, the field equations of a theory can be grounded on the relevant physical phenomena with a comfortable degree of certainty whereas possible boundary conditions may not be known or understood with such certainty. In such a case, the device of exploring conditions admitted at the boundary that lead to formal uniqueness of solutions to the field equations, should they exist, becomes a useful tool. The field equations may be nonlinear and consist of coupled systems of equations, so that deriving even a formal uniqueness theorem for the entire system may not be a straightforward proposition. In such cases, deriving a formal uniqueness theorem for a suitably simplified version of the theory, e.g. breaking the system into parts and deriving formal uniqueness conditions for each such part by assuming some fields as imposed data, may be an accessible option, as has been done in Acharya (2003). Given the complexity of our mesoscopic model when appended with suitable constitutive equations, it is the latter option that we choose to exercise for inferring plausible boundary conditions for the model.

Equation $(1)_{1,2}$, now interpreted as averaged equations, augmented with the condition

$$\bar{\boldsymbol{\chi}} \boldsymbol{n} = \boldsymbol{0} \text{ on the boundary } \partial B \text{ of the body with outward unit normal } \boldsymbol{n}, \tag{15}$$

are sufficient for uniqueness, when the field $\bar{\boldsymbol{\alpha}}$ is treated as data. Physically, all we want to impose is the statement of elastic incompatibility in the interior of the body and we would like to be able to solve such an equation in a stable fashion. The requirement that $\bar{\boldsymbol{\chi}} \boldsymbol{n}$ vanish on the boundary is because we would like the 'normal action' of $\bar{U}^e$ on the boundary to be determined solely from the slipping distortion. Equation (5) admits the condition

$$\dot{\bar{z}} \text{ specified at an arbitrarily chosen point of the body} \qquad (16)$$

for uniqueness of the field $grad\,\dot{\bar{z}}$, when $S$ is treated as data. Equation $(1)_4$ admits the usual variety of essential and/or natural boundary conditions on displacements and tractions, when $grad\,\bar{z}$ and $\bar{\chi}$ are treated as data. This leaves (6) as the only outstanding equation requiring attention with respect to boundary conditions.

To begin, we note that (6) is an equation for evolution of dislocation density that is similar to the local statement of conservation for an areal density of (essentially) lines (Mura, 1963; Acharya, 2001). However, this local statement can also be integrated over the volume of the body yielding a statement that appears to have the form of a volumetric conservation law for $\bar{\alpha}$, with associated natural boundary conditions. Since we are interested in solving (6) in finite regions of three-dimensional space, this is important for us. Following Nye (1953), $\bar{\alpha}$ is most naturally viewed as an areal density whose action on a unit direction yields the Burgers vector per unit area of all dislocation lines piercing an area perpendicular to the unit direction. To facilitate a volumetric interpretation, $\bar{\alpha}$ may also be viewed as a volumetric density as follows (Arsenlis and Parks, 1999; Acharya, 2003): the action of $\bar{\alpha}$ on a unit direction is to yield the sum of the products of the Burgers vector and line length along the unit direction, per unit volume, of all the dislocation lines present in that volume. Of course, we rest content in the knowledge that we are working on a three-dimensional body in three-dimensional Euclidean space so that tensors defined on tangent spaces of distinct points may be added with impunity, and integration of a tensor-valued volume density makes sense.

2.2.1 *The Surface Flow and its physical interpretation*

It is a straightforward matter to multiply (6) by a test function and then integrate over the body (Acharya, 2003) or, alternatively, directly integrate over the body or a part (Gurtin and Needleman, 2005) to uncover a surface flow term on a bounding surface of a part/body with unit normal field $n$ as

$$S \times n\,.^{iv} \qquad (17)$$

---

[iv] An important, essential difference of our model with that of Gurtin and Needleman (2004) is that the field equation (6) from which we derive the dislocation boundary conditions is not an identity in the sense that the equation is not identically satisfied by mere virtue o f the fact that the rest of the governing equations of the model have been solved.

The surface flow (17) may be interpreted as follows: at any point of the boundary consider an orthonormal triad $(\boldsymbol{e}_1, \boldsymbol{e}_2, \boldsymbol{e}_3)$ with $\boldsymbol{e}_3 = \boldsymbol{n}$. At this point the flow may be written as

$$\boldsymbol{S} \times \boldsymbol{n} = (\boldsymbol{S}\boldsymbol{e}_i) \otimes (\boldsymbol{e}_i \times \boldsymbol{n}) = (\boldsymbol{S}\boldsymbol{e}_2) \otimes \boldsymbol{e}_1 - (\boldsymbol{S}\boldsymbol{e}_1) \otimes \boldsymbol{e}_2. \tag{18}$$

Thus, in terms of matrix representations on the $(\boldsymbol{e}_1, \boldsymbol{e}_2, \boldsymbol{e}_3)$ basis, the surface flow consists of the matrix whose first column is the second column of the slipping distortion, its second column is the negative of the first column of the slipping distortion with its third column vanishing identically. Importantly, this decomposition implies that vanishing of the surface flow does not pose any constraint on slipping distortion components $S_{i3}$, with respect to the above basis, at the boundary; in particular, slipping distortions characterizing simple shears on the boundary tangent plane can take place with impunity with such a no-flow boundary condition in force, i.e. *a no-flow boundary condition does not imply a boundary layer of low shear strain in simple shear*. Physically, this is reasonable as such simple shears are associated with boundary-parallel motion of dislocation lines parallel to the boundary (where we think of the macroscopic 'boundary' as a layer of a few atoms within which the dislocation line exists), and unless there are geometric constraints (e.g. unavailable slip system) preventing such slipping, there does not appear any reason for the dislocation line not to be able to move the 'bulk' material to accomplish this slipping. Of course, such unrestricted motions can also result in normal strains – for instance, climb of an edge dislocation parallel to the boundary, where the burgers vector of the dislocation is normal to the boundary.

*2.2.2 Connection between the Surface Flow and Flux*

Given the existence of a kinematical connection between dislocation motion and slipping distortions as the ones just invoked, it is perhaps natural to ask as to what type of a *flux* expression may be related to the motion of dislocation lines through the surface and what the connection of such a flux may be to the surface flow (17). By a flux we simply mean the entry/exit of infinitesimal dislocation segments into/out of the body, carrying with them their associated Burgers vector. Thus, consider a dislocation distribution at a boundary point of the form $\boldsymbol{b} \otimes \boldsymbol{l}$ moving with a velocity $\bar{\boldsymbol{V}}$ parallel to the boundary. The surface flow now takes the form $\boldsymbol{b} \otimes \{(\boldsymbol{l} \times \bar{\boldsymbol{V}}) \times \boldsymbol{n}\}$. Regardless of the orientations of $\boldsymbol{b}$ and $\boldsymbol{l}$, such a motion should not correspond to a flux of the dislocation distribution through the surface; however, unless $\boldsymbol{l}$ is parallel to the boundary the surface flow (17) in this case is non-zero. This suggests that for slipping distortions that can be represented as the 'cross-product' of a dislocation density tensor and a velocity vector, the surface flow (17) is not a geometrically accurate measure of flux of dislocation density into (out of) the body. An accurate measure of a flux of dislocation density in such a case may be considered to be the term

$$\bar{\alpha}(\bar{V} \cdot n) \tag{19}$$

which is an additive component in the expression $(\bar{\alpha} \times \bar{V}) \times n$, as can be seen from the formula for the non-commutative vector product of three vectors. Indeed, the less-restrictive specification of the dislocation flux expression (19) on inflow parts of the boundary (boundary points where $\bar{V} \cdot n < 0$) is sufficient for uniqueness of solutions to (6) in simplified situations, e.g. when $L^p$, $\bar{V}$ and $\bar{s}$ are assumed to be specified functions of space and time, as can be seen by applying verbatim the arguments in Acharya (2003) to this context.

2.2.3 *Imposition of physically motivated boundary conditions on Surface Flow*

With the geometric characterization of the surface flow and flux in hand, we now ask as to what conditions may be imposed on such quantities to qualify as boundary conditions for (6). Purely on physical grounds, a *zero flow boundary condition*, $S \times n = 0$, on the entire boundary would seem to be appropriate in many circumstances in order to represent a rigid boundary with regard to slipping, as can be seen from the prohibited slipping distortion modes following the discussion surrounding (18). Of course, such a condition can lead to the development of shocks, or discontinuities, in the dislocation density field as the no-flow constraint may be operative on outflow boundaries restricting an outward flux of dislocations. Due to the presence of the term $L^p$ in $S$, a natural boundary condition of the form, $S \times n = \Phi$, where $\Phi$ is a specified function of time and position along the boundary satisfying the constraint $\Phi n = 0$, may also be appropriate to model controlled flow at the boundary. A less-restrictive condition is the imposition only of $\bar{\alpha}(\bar{V} \cdot n)$ on inflow points of the boundary (i.e. boundary points where $\bar{V} \cdot n < 0$) along with a specification of $L^p \times n$ on the entire boundary. This condition allows free exit of GNDs at outflow boundary points without any added specification, as shown in the formation of slip steps in Roy and Acharya (2005), and this is different from the condition suggested by Arsenlis et al. (2004) to achieve similar effects. The boundary condition specifications above seem to cover situations that we can think of as being relevant in applications at the present time.

2.2.4 *Contact with the work of Gurtin and Needleman (2005)*

We end this section by making contact with the work of Gurtin and Needleman (2005). If their arguments are applied to (6) with $\bar{s} = 0$, then a result of theirs suggests that "no flow of Burgers vector across" the boundary surface element is *equivalent* to $S \times n = 0$ whereas, for us, the physical idea of no flux of existing dislocation lines through the surface does not necessarily imply $S \times n = 0$, as shown earlier. The apparent discrepancy appears to be a matter of definition, as shown next.

Gurtin and Needleman show that at a boundary point with normal $n$, given any unit vector $e$, the flux of Burgers vector associated with circuits perpendicular to $e$ is given by

$$-(S \times e)n = (S \times n)e. \tag{20}$$

Since $e$ is arbitrary, we choose it to be $n$ and find from (20) that the corresponding flux of Burgers vector vanishes, regardless of $S$. However, it is of course possible to conceive of a tensor $S$ such that $S \times n \neq 0$. Clearly, then, Gurtin and Needleman's definition of no flux of Burgers vector across the boundary amounts to demanding

$$(S \times e)n = 0 \ \text{for all} \ e, \tag{21}$$

regardless of whether it is physically important at all to consider the flux of Burgers vector associated with circuits perpendicular to directions parallel to which no dislocation lines may be present at the boundary.

Interestingly, if one is willing/forced to be ambiguous about relating the slipping distortion to the motion of dislocation lines and their orientation, then the flow (17) can always be identified with a flux of dislocations. In terms of the orthonormal triad introduced just before (18), the slipping distortion may be written as

$$S = (Se_1) \otimes e_1 + (Se_2) \otimes e_2 + (Se_3) \otimes e_3. \tag{22}$$

We have seen that the flow (17) involves only the first two slipping modes on the right hand side of (22). Now, up to signs, a slipping distortion like $(Se_1) \otimes e_1$ can be effected by (vector components of) dislocation lines normal to the boundary moving in the $e_2$ direction, in which case there is no flux of dislocations through the boundary, or by dislocation lines in the $e_2$ direction moving in the $e_3 = n$ direction, in which case there is a flux of dislocations. This situation is analogous to identical simple shearing produced by motion on the same slip plane of edge or screw dislocations with common Burgers vector. Consequently, a non-zero flow (17) can always be related to a nonzero flux of dislocations if the 'state of dislocations' at the boundary is assumed to be assignable freely. Indeed, even in our case, if $\bar{V} \equiv 0$, then $S \equiv L^p$ is the slipping distortion produced by SSDs and, by definition, this cannot be attributed to the motion of a dislocation density unambiguously. However, since we have an additional component $(\bar{\alpha} \times \bar{V})$ in the slipping distortion $S$ that is associated with the state of GNDs at a material point, a no-flow condition cannot be considered as equivalent to a no-flux condition.

Indeed, it is precisely for such reasons that the flux (19) was uncovered in Acharya (2003), ignoring the flow term $(\bar{\alpha} \times \bar{V}) \times n$ - that results from a direct integration by parts of the evolution equation for dislocation density - as a boundary condition in that context.

*2.2.5 Initial conditions for averaged field equations*

The averaged field equations admit initial conditions on the fields $\bar{u}$, $\bar{\alpha}$, and $grad\,\bar{z}$. We assume that a physically natural initial condition on the displacement field is $\bar{u}|_{t=0} \equiv 0$. Given the initial condition for the field $\bar{\alpha}$, the initial condition on $grad\,\bar{z}$ is determined by solving the averaged equations corresponding to $(1)_{1,2,4}$ and (15) with $\bar{u}|_{t=0} \equiv 0$ for the field $\bar{z}|_{t=0}$, with the latter's value specified at one single point of the body arbitrarily. This problem also determines the internal stress field in the body at the initial instant corresponding to the dislocation density distribution $\bar{\alpha}|_{t=0}$.

## 3. A specific model of PMFDM, contrast of an idea with its counterpart in classical Continuously Distributed Dislocation theory, and connections to Lower-order Gradient Plasticity

From here onwards, *fields without overhead bars refer to averaged fields*. Sometimes, we retain the overhead bars to avoid confusion. If a field without overhead bars is used to represent a microscopic field, we mention this specifically.

Motivated by the considerations of the preceding section, the governing field equations of the model of mesoscopic plasticity we propose are

$$curl\,\chi = \alpha$$
$$div\,\chi = 0$$
$$\chi n = 0 \text{ on boundary of body}$$
$$div(grad\,\dot{z}) = div(\alpha \times V + L^p)$$
$$(grad\,\dot{z})n = (\alpha \times V + L^p)n \text{ on boundary of body} \quad (23)$$
$$\dot{z} \text{ specified arbitrarily at one point of body}$$
$$U^e = grad(u - z) + \chi$$
$$div\,T = 0$$
$$\dot{\alpha} = -curl(\alpha \times V + L^p)$$

where the stress tensor $T$, the dislocation velocity vector $V$, and the slip distortion rate produced by SSDs, $L^p$, are to be specified constitutively. Here, we make the modeling choice $\bar{s} \equiv 0$.

The (symmetric) stress tensor is specified according to linear elasticity as

$$\boldsymbol{T} = \boldsymbol{C} : \boldsymbol{U}^e, \qquad (24)$$

where $\boldsymbol{C}$ is the spatially uniform, fourth-order tensor of elastic moduli with major and minor symmetries. We would like to ensure that the plastic working be non-negative by our constitutive choices so as to qualify as dissipative effects, as well as guarantee pressure-insensitivity of plasticity at the mesoscale (recognizing that the latter constraint may very well not be physically appropriate after detailed study in the future). Simple choices that satisfy the above requirements have the form

$$\begin{aligned}
\boldsymbol{L}^p &= \dot{\gamma} \frac{\boldsymbol{T}'}{|\boldsymbol{T}'|} \quad ; \quad \dot{\gamma} \geq 0 \\
\boldsymbol{V} &= v \frac{\boldsymbol{d}}{|\boldsymbol{d}|} \quad ; \quad v \geq 0, \\
\boldsymbol{b} &:= \boldsymbol{X}(\boldsymbol{T}'\boldsymbol{\alpha}) \; ; \; b_i = e_{ijk} T'_{jr} \alpha_{rk} \; ; \; \boldsymbol{a} := \boldsymbol{X}\left(\frac{1}{3} tr(\boldsymbol{T})\boldsymbol{\alpha}\right) \; ; \; a_i = \left(\frac{1}{3} T_{mm}\right) e_{ijk} \alpha_{jk} \\
\boldsymbol{d} &:= \boldsymbol{b} - \left(\boldsymbol{b} \cdot \frac{\boldsymbol{a}}{|\boldsymbol{a}|}\right) \frac{\boldsymbol{a}}{|\boldsymbol{a}|}
\end{aligned} \qquad (25)$$

where $\boldsymbol{T}'$ is the stress deviator and $\dot{\gamma}$ and $v$ are non-negative functions of state representing the magnitudes of the SSD slipping rate and the averaged dislocation velocity.

For the purpose of formulating constitutive relations for $\dot{\gamma}$ and $v$, let $\rho_m$ be the line-length per unit volume of mobile dislocations within the averaging volume and assume it to be some fraction, $f$, of the obstacle density $\rho_o$. From the Orowan and the empirical Taylor/Bailey-Hirsh relationships

$$\begin{aligned}
\dot{\gamma} &= \rho_m b v \\
g &= \eta \mu b \sqrt{\rho_o}
\end{aligned} \qquad (26)$$

where $g$ is the strength of the material, $\eta \approx 1/3$, $\mu$ is the shear modulus, and $b$ is the magnitude of the Burgers vector of the material. Eliminating $\rho_m = f \rho_0$ from (26) allows expressing $v$ in terms of $\dot{\gamma}$, and we assume the latter to be known from experiments as a function of flow stress and strength, e.g. standard power law relationships. Thus, making the further gross assumption that $f = 1$,

$$v(state) = \eta^2 b \left(\frac{\mu}{g}\right)^2 \dot{\gamma}(\boldsymbol{T}', g). \qquad (27)$$

In this work we shall choose the power law representation

$$\dot{\gamma} = \dot{\gamma}_0 \left( \frac{|\boldsymbol{T'}|}{\sqrt{2}g} \right)^{\frac{1}{m}}, \qquad (28)$$

where $m$ is the macroscopic rate-sensitivity of the material and $\dot{\gamma}_0$ is a reference strain rate (complete with the usual abuse of notation).

Finally, we need an evolution equation for the flow strength of the material, $g$, and we utilize the equation proposed by Acharya and Beaudoin (2000) with obvious modifications for the hardening due to GNDs and the operative plastic strain rate:

$$\dot{g} = \left[ \frac{\eta^2 \mu^2 b}{2(g - g_0)} k_0 |\boldsymbol{\alpha}| + \theta_0 \left( \frac{g_s - g}{g_s - g_0} \right) \right] \{|\boldsymbol{\alpha} \times \boldsymbol{V}| + \dot{\gamma}\}. \qquad (29)$$

In the above, $g_s$ is a saturation stress, $g_0$ is the yield stress, and $\theta_0 \approx \mu/200$ is the Stage II hardening rate. The material parameters $g_s, g_0, \mu, b, \dot{\gamma}_0, m$ are known from conventional plasticity (Voce Law). Consequently, $k_0$ is the only extra parameter that needs to be fitted and can be obtained from experimental grain-size dependence of flow stress results, as shown in Acharya and Beaudoin (2000) and Beaudoin et al. (2000).

### *3.1 Slipping Distortion, Plastic part of Velocity Gradient and a difference of interpretation from the classical Continuously Distributed Dislocations theory*

*If as much smoothness as required is assumed of all fields*, then it can be shown that

$$\dot{\boldsymbol{U}}^p = \boldsymbol{\alpha} \times \boldsymbol{V} + \boldsymbol{L}^p = \boldsymbol{S} \qquad (30)$$

as follows. Equations (8) and (23)$_{1,9}$ imply

$$\operatorname{curl} \dot{\boldsymbol{U}}^p = \operatorname{curl}(\boldsymbol{\alpha} \times \boldsymbol{V} + \boldsymbol{L}^p), \qquad (31)$$

while (8) and (23)$_{2,4}$ imply

$$\operatorname{div} \dot{\boldsymbol{U}}^p = \operatorname{div}(\boldsymbol{\alpha} \times \boldsymbol{V} + \boldsymbol{L}^p). \qquad (32)$$

Finally, (31), (32), (8), and (23)$_{3,5}$ imply (30). A model based on (30) as a fundamental equation has the advantage that (23)$_{1-7}$ become redundant equations and one (almost) obtains the structure of conventional plasticity with the plastic distortion rate given by (30). Up to the form of the plastic distortion rate, the above argument, in essence, was stated in Mura (1963, 1970), and provides a strong conceptual link between the continuum theory of dislocations and plasticity theory. Of course, the residual/initial stress state corresponding to the initial dislocation density distribution can only be computed with the help of some of the eliminated equations (Acharya, 2001; Roy and Acharya, 2005).

We now examine the abovementioned reduction in some detail. The governing equation for $\alpha$ admits spatial discontinuities in the field and, along with the effect of the boundary conditions, also on the corresponding (tensor) $S$. Moreover, the constitutive equations for $V$ and $L^p$ depend on the stress tensor subject to only traction continuity on arbitrary internal surfaces. Importantly, it is a fundamental result of ECDD (Kröner, 1981) that gradients in $\alpha$ induce long-range internal stresses, and hence it is our opinion that $\chi$ should at least be a spatially continuous field to sense and instantaneously communicate long-range effects. Thus, on a heuristic basis, it seems to us that the spatial smoothness of the incompatible parts of the tensor fields $\dot{U}^p$ and $S$ should be different, and we do not consider these fields as identical. Said differently, we interpret the solution of $(23)_{1-3}$ as also satisfying the following second-order, Poisson's equation

$$div(grad\ \chi) = -curl\ \alpha \qquad (33)$$

in keeping with the interpretation of ECDD, due to the divergence-free condition on $\chi$. However, no part of the slipping distortion field $S$ is required to satisfy a second-order, elliptic, differential equation. Thus, when interpreted in the weak sense, the field $\chi$ is required to have at least square-integrable derivatives whereas it suffices for $S$ itself to be simply square-integrable. Interpreted yet another way, if the formula $-curl\ U^p$ were to be substituted in (30) for $\alpha$ utilizing (8) (typically $V$ would also have to be a function of $\alpha$), then (30) would be a genuinely nonlinear, first-order system in $U^p$ requiring low regularity on the field for any reasonable interpretation of solutions to the system, and such regularity would be at odds with that required of $U^p$ for an adequate prediction of internal stress effects through the equilibrium equation $div\{C:(grad\ u - U^p)\} = 0$ with a continuous displacement field[v].

The above, along with the fact that the compatible part, $grad\ \dot{z}$, of $\dot{U}^p$ depends non-locally on $S$ ($(23)_{4-5}$), implies that the pointwise values of $\dot{U}^p$ and $S$ in our model need bear no resemblance to each other. The importance of these features is demonstrated in Part II of this paper through the prediction of back stress in strict simple shearing with constitutive equations of the form (27)-(28), i. e. *with nothing more than the elastic stress in the driving force for slip* (cf. Lemaitre and Chaboche, 1990; Menzel and Steinmann, 2000; Gurtin, 2002; Arsenlis et al., 2004; Yefimov et al. 2004), which is also in the spirit of Discrete Dislocation Plasticity (e.g. Benzerga et al., 2005). Henceforth, we refer to

$$\dot{U}^p = grad\,\dot{z} - \dot{\chi} \qquad (34)$$

as the *plastic part of the velocity gradient.*

---

[v] We note here that inasmuch as the total displacement of the theory does not represent the actual physical motion of atoms involving topological changes but only a consistent shape change, it is not required to be discontinuous. However, the stress produced by these topological changes is adequately reflected in the theory through the utilization of incompatible elastic/plastic distortions.

Based on the considerations above, our specification of the plastic distortion $U^p$ may be viewed as a physically motivated, non-local prescription of a 'transformation distortion' field (Eshelby, 1956, 1957). The incompatible part of this field is driven by the instantaneous dislocation density field with its compatible part being dependent on the history of slipping due to GND and SSD populations in the material.

### *3.2 Lower-order Gradient Plasticity and associated Boundary Conditions*

Lower-order gradient plasticity is a modification of conventional plasticity theory capable of predicting size-effects, where a GND measure enters only into the hardening modulus (Acharya and Bassani, 1996; 2000). Such a modification has some desirable properties with respect to the definition of the boundary-value-problem of incremental equilibrium (Acharya and Shawki, 1995, Acharya and Bassani, 1996), especially in comparison to other higher-order gradient plasticity proposals (Fleck and Hutchinson, 2001, Gurtin, 2000, Huang et al., 2000). It has recently been shown in a one-dimensional setting that lower-order gradient plasticity can admit boundary conditions (Acharya et al. 2004), despite its 'lower-order' structure. However, the structure of plausible boundary conditions in the three-dimensional case did not become clear in the aforementioned study. Interestingly, the present work recovers lower-order gradient plasticity trivially as a constitutive limit and suggests a plausible boundary condition on plastic strain rates that may be applied to model constrained plastic flow.

*Enhanced lower-order gradient plasticity* is recovered by setting $V \equiv 0$ in the set of equations (23), (24), $(25)_1$, (28), and (29). Long-range stress effects due to gradients in dislocation density arising solely from SSD evolution are expected to be recovered by this approximation as well as enhanced work-hardening due to GND content.

*Standard lower-order gradient plasticity* is recovered through the set $(23)_{8,9}$, (24), $(25)_1$, (28), (29), and (30) with $V \equiv 0$ operative and $U^e := grad\,u - U^p$. Work-hardening due to GNDs is modeled, but possible long-range stress effects due to appropriate gradients in dislocation density are not expected to be recovered adequately, even when such gradients exist in the dislocation density field for reasons mentioned in Section 3.1.

In both cases, the boundary condition

$$L^p \times n = \Phi \qquad (35)$$

may be specified, where $\Phi$ is a function of time and position on the boundary, satisfying the constraint $\Phi n = 0$, i.e. only the 'tangential action' of the plastic part of the velocity gradient at the boundary is specified.

## 4. Power Expenditures, Work hardening

In the absence of inertia, the mechanical power expended by all external agencies (traction and displacement boundary conditions, body forces) on the body is partitioned in the model as

$$\int_B \boldsymbol{T} : grad\, \dot{\boldsymbol{u}}\, dv = \int_B \dot{\Psi}\, dv + \int_B \left[ v|\boldsymbol{d}| + \dot{\gamma}|\boldsymbol{T}'| \right] dv$$
$$\Psi(\boldsymbol{U}^e) := \frac{1}{2} \boldsymbol{U}^e : \boldsymbol{C}\boldsymbol{U}^e, \tag{36}$$

as follows from (8), (9), (11), (24), and (25). In (36), the first integral on the right hand side *may be* interpreted as the rate of change of stored elastic energy and the second term as the dissipation implied by the model due to the latter's non-negativity.

### 4.1 *On the need for a free-energy function in mesoscopic dislocation mechanics*

At the microscopic level, dislocations induce added stored energy through lattice stretching. Consequently, we believe that standard descriptions of elasticity should be adequate to account for the added stored energy of dislocation distributions in any model, but that the description of elastic strains in the model has to be enhanced to account for the energetic effects of defects. Thus, assuming a completely resolved dislocation distribution (for dislocations with spread out cores, a physically realistic feature), the elastic theory of continuously distributed dislocations and FDM (Kröner, 1981; Roy and Acharya, 2005) calculate the strain energy of the dislocation distribution exactly with *a strain energy dependence on only the elastic strain*. For that matter, if the instantaneous plastic distortion in conventional plasticity is such that the field resulting from taking its *curl* is identical to a distribution of discrete dislocations (with spread out cores) – such a situation can always be arranged as a specific initial condition on plastic strain – then the stress field and elastic strain energy predicted by conventional plasticity would also be exactly correct. Thus, it is the prescription of plastic straining in conventional plasticity that needs to be enhanced to account for the presence of dislocations, and this is one of the express goals of FDM and PMFDM.

Of course, PMFDM is meant to be a model of dislocation plasticity at scales of resolution much coarser than where every dislocation is well-resolved, so the appropriate form of a stored energy function for such a situation becomes an issue. But even here, if the coarse free energy function is meant to represent the spatially averaged free energy content in a representative volume element for the coarse-scale model, then an added dependence on the average dislocation density tensor cannot be adequate. For it is well understood that for two different spatial distributions of microscopic dislocation density $\boldsymbol{\alpha}_1$ and $\boldsymbol{\alpha}_2$ within the representative volume with identical average $\bar{\boldsymbol{\alpha}}_1 = \bar{\boldsymbol{\alpha}}_2$, the averaged free-energy content is, in general, different, i.e.

$\overline{\psi(\boldsymbol{\alpha}_1)} \neq \overline{\psi(\boldsymbol{\alpha}_2)}$, where $\psi$ represents the microscopic (specific) free energy function. But a model that proposes to account for the added strain energy of dislocations at the coarse-scale by a unique functional dependence on $\bar{\boldsymbol{\alpha}}$, say $R(\bar{\boldsymbol{\alpha}})$, alone is bound to fail since the evaluation of such a function for the two cases would have to be identical:

$$R(\bar{\boldsymbol{\alpha}}_1) = R(\bar{\boldsymbol{\alpha}}_2) \text{ but } \overline{\psi(\boldsymbol{\alpha}_1)} \neq \overline{\psi(\boldsymbol{\alpha}_2)}. \tag{37}$$

Therefore, it seems to us that the various models proposed in the literature to account for the energetic effects of dislocations at the mesoscopic scale by an added dependence of the free energy on the average dislocation density tensor actually represent something other than the spatial (running) average of the microscopic free-energy content of the body by this preferred device.

Fortunately, the mechanical structure of PMFDM arises from an averaging of the microscopic equations *without any need for a mesoscopic free energy function*, and even in the constitutive structure – only related to $V, s, L^p$ with the average stress entering the equilibrium equation completely determined - no phenomenology related to 'back-stress' need necessarily be introduced.

### 4.2 *The dissipative stress*

In order to discuss energy balance including the effects of heat and temperature at the mesoscale, the first law of thermodynamics at the microscopic (FDM) level can be averaged with the result

$$\dot{\bar{\varepsilon}} = \bar{\boldsymbol{T}} : grad\,\dot{\bar{\boldsymbol{u}}} + div\,\bar{\boldsymbol{q}} + \overline{(\boldsymbol{T} - \bar{\boldsymbol{T}}):(grad_{x'}\dot{\boldsymbol{u}} - grad\,\dot{\bar{\boldsymbol{u}}})}, \tag{38}$$

where $\varepsilon$ is the microscopic internal energy per unit volume and $\boldsymbol{q}$ is the microscopic heat flux vector defined by $\boldsymbol{q} \cdot \boldsymbol{n}$ yielding the (scalar) heat flux into the body through a surface element with outward unit normal $\boldsymbol{n}$. Restricting attention to isothermal situations at averaged temperature $\theta_0$ for simplicity, introducing a macroscopic free energy function $\Psi$ as in (36) along with the notions of macroscopic temperature $\theta$ and entropy $\eta$ as in $\bar{\varepsilon} = \Psi - \eta\theta$, and *ignoring the fluctuation stress-power term in* (38), we arrive at

$$\int_{\partial B} -\boldsymbol{q} \cdot \boldsymbol{n}\,da = \int_{B} \left[ v|\boldsymbol{d}| + \dot{\gamma}|\boldsymbol{T}'| \right] dv - \int_{B} \theta_0 \dot{\eta}\,dv, \tag{39}$$

where we have dropped overhead bars in (39) for convenience. Granted this approximate analysis, the mechanical dissipation due to plasticity, less any entropy changes in the body due to plastic deformation, is converted to heat given out by the body. It is of course tempting to speculate that the fluctuation stress-power term in (38) need not be ignored in deriving (39) and actually constitutes the entropic term so that the plastic dissipation is, in entirety, a source for the average heat flux out of the body (plus the average temperature rise in case of non-isothermal situations).

An experimental result of plasticity at the macroscopic scale due to Taylor and Quinney (1934) suggests that the ratio of the rate at which heat is given out to the rate of mechanical working during plastic deformation at room temperature follows a nominally increasing trend. This experimental result implies that the *dissipative 'stress'* in models, power-conjugate to the rate of plastic straining in the expression for stress power (e.g. (36)), *has to increase with the observed strength of the material* as measured, say, on an experimental stress-strain curve under monotonic loading, assuming that the rate of elastic straining can be ignored at such scales. The more recent results of Rosakis et al. (2000) are nominally consistent with this trend, while differing in detail. Such a feature is in contrast to some thermodynamic models of plasticity in the literature, conventional as well as of incremental deformation theory type, where the dissipative power-conjugate stress remains constant or vanishes.

The constitutive structure of PMFDM relies heavily on the idea of work hardening leading to increased dissipation for otherwise equal plastic strain increments. This is evident from the term representing dissipation in the model,

$$\int_B \dot{\gamma} \left( \eta^2 b \mu^2 g^{-2} |d| + |T'| \right) dv , \qquad (40)$$

coupled with the constitutive equation (28), especially in the rate-insensitive case. In (40), the sum within the parenthesis represents the dissipative stress in the model. Interestingly, it depends upon the GND density $\alpha$ and a material length scale given by $\eta^2 (\mu/g)^2 b$, even though only *increments* in strength depend on $\alpha$ (29).

**4.3** *On the possibility of a back-stress tensor in PMFDM*

We mention here in passing that *if additional thermodynamic phenomenology at the mesoscale is introduced,* PMFDM does not preclude the existence of an entity in the constitutive structure that might provide an effect much like a back-stress as in conventional plasticity, as we show in the following paragraphs of this section. However, as can be seen from the discussion surrounding (38), a strict interpretation of energy balance on averaging appears to add more indefiniteness than clarification. We consider it a strength of our approach that such additional thermodynamic postulation is not required, and yet we are able to predict strong Bauschinger effects without the introduction of any back-stress-like entity in our constitutive model (Part II). Consequently, for the present, we invoke Occam's Razor as a guiding principle in phenomenology in not presuming that what is not required.

To show how a back-stress-like entity might appear with added thermodynamic phenomenology, for the microscopic free energy density $\psi = 1/2(U^e : CU^e)$ (in terms of the microscopic elastic distortion), consider

$$\bar{\psi}(x,t) = \frac{1}{2}\bar{U}^e(x,t) : C\bar{U}^e(x,t) + \frac{1}{2}\overline{(U^e - \bar{U}^e(x,t)) : C(U^e - \bar{U}^e(x,t))}, \tag{41}$$

where the cross-terms in the fluctuations and the mean fields vanish on averaging.

The fluctuation in the elastic distortion cannot be written *uniquely* in terms of the instantaneous values of the mean fields, in general. Within a phenomenological approach, this dictates the introduction of an additional fourth-order tensor-valued state variable

$$\boldsymbol{\beta}(x,t) := \frac{1}{2}\overline{(U^e - \bar{U}^e(x,t)) \otimes (U^e - \bar{U}^e(x,t))} \tag{42}$$

that is physically dimensionless, so that

$$\bar{\psi}(\bar{U}^e, \boldsymbol{\beta}) = \frac{1}{2}\bar{U}^e : C\bar{U}^e + C \cdot^{(4)} \boldsymbol{\beta}$$

$$C \cdot^{(4)} \boldsymbol{\beta} := C_{ijkl}\beta_{ijkl}, \tag{43}$$

where $\cdot^{(4)}$ represents the inner product of two fourth order tensors. In (42) the unbarred quantities are microscopic fields. If we now consider the function $\bar{\psi}$ as the mesoscopic free energy density, then the mesoscopic dissipation for the mechanical model described by (23) and (24) is expressed as

$$\int_B \bar{T} : grad\,\dot{\bar{u}}\,dv - \int_B \dot{\bar{\psi}}\,dv = \int_B \bar{T} : L^p\,dv + \int_B \boldsymbol{\xi} \cdot \bar{V}\,dv - \int_B C \cdot^{(4)} \dot{\boldsymbol{\beta}}\,dv. \tag{44}$$

A constitutive equation is required for the evolution of $\boldsymbol{\beta}$. To make contact with conventional plasticity, consider the hypothetical prescription

$$\dot{\boldsymbol{\beta}} := cL^p \otimes \boldsymbol{\Delta}, \tag{45}$$

where $\boldsymbol{\Delta}$ is second order distortion tensor with magnitude of the order of elastic strain and $c$ is a suitably small, positive, dimensionless constant so that

$$\bar{T} - cC\boldsymbol{\Delta} \tag{46}$$

is the dissipative power-conjugate to $L^p$ in (44), and

$$cC\boldsymbol{\Delta} \tag{47}$$

may be considered as a 'back-stress' tensor.

Of course, in reality there is no reason to expect $\boldsymbol{\beta}$ or $\dot{\boldsymbol{\beta}}$ to be the dyadic product of two second-order tensors and the constitutive specification of $\dot{\boldsymbol{\beta}}$ is not a simple matter.

## 5. Finite deformation, Single Crystal Plasticity

It is well understood that finite lattice rotations and nonlinear crystal elasticity have observable effects in the meso/macroscopic plastic response of single and polycrystalline materials. Motivated by the considerations of the previous sections and the theory presented in Acharya (2004) for finite deformation Field Dislocation Mechanics, we propose the following model of finite deformation, mesoscopic single crystal plasticity:

$$\begin{aligned}
&\text{curl } \tilde{\boldsymbol{\chi}} = -\boldsymbol{\alpha} \\
&\text{div } \tilde{\boldsymbol{\chi}} = \mathbf{0} \\
&\text{div}\left(\text{grad } \dot{\boldsymbol{f}}\right) = \text{div}\left(\boldsymbol{\Pi} + \tilde{\boldsymbol{L}}^p - \dot{\tilde{\boldsymbol{\chi}}} - \tilde{\boldsymbol{\chi}} \boldsymbol{L}\right) \\
&\boldsymbol{F}^{e-1} = \tilde{\boldsymbol{\chi}} + \text{grad } \boldsymbol{f} \quad ; \quad \boldsymbol{C}^e = \boldsymbol{F}^{eT} \boldsymbol{F}^e \\
&\boldsymbol{T} = 2\rho \boldsymbol{F}^e \frac{\partial \psi}{\partial \boldsymbol{C}^e} \boldsymbol{F}^{eT} \\
&\text{div } \boldsymbol{T} = \mathbf{0} \\
&(\text{div } \boldsymbol{v})\boldsymbol{\alpha} + \dot{\boldsymbol{\alpha}} - \boldsymbol{\alpha} \boldsymbol{L}^T \equiv \overset{\circ}{\boldsymbol{\alpha}} = -\text{curl}\left(\boldsymbol{\Pi} + \tilde{\boldsymbol{L}}^p\right).
\end{aligned} \qquad (48)$$

In (48), all spatial derivative operators are with respect to the current configuration and superposed dots represent material time derivatives. The field $\boldsymbol{v}$ represents the velocity field with $\boldsymbol{L} = \text{grad } \boldsymbol{v}$ the velocity gradient. $\psi$ is the free energy per unit mass dependent only on $\boldsymbol{C}^e$ and $\rho$ is the density. $\boldsymbol{\alpha}$ represents the two-point tensor of dislocation density between the current and the unstretched lattice configuration and $\boldsymbol{F}^e$ is the elastic distortion tensor. Given slip systems indexed by $\kappa$ with unstretched unit slip direction $\boldsymbol{m}_0^\kappa$ and unit slip plane normal $\boldsymbol{n}_0^\kappa$, the two-point SSD slipping distortion is given by

$$\tilde{\boldsymbol{L}}^p = \sum_\kappa \dot{\gamma}^\kappa \boldsymbol{m}_0^\kappa \otimes \boldsymbol{n}_0^\kappa \boldsymbol{F}^{e-1}, \qquad (49)$$

where we use notation in common use even though the scalar slippings $(\dot{\gamma}s)$ are not actual material time derivatives of any quantity in our model.

Two alternatives for the GND slipping distortion are proposed. The first alternative prescribes

$$\boldsymbol{\Pi} := \boldsymbol{\alpha} \times \boldsymbol{V}. \qquad (50)$$

In this case, the dissipation in the model can be shown to be

$$\int_C \boldsymbol{X}(\boldsymbol{T}\boldsymbol{\alpha}^*) \cdot \boldsymbol{V} \, dv + \int_C \sum_\kappa \tau^\kappa \dot{\gamma}^\kappa \, dv \qquad (51)$$

where $C$ is the current configuration,

$$\alpha^* = F^e \alpha \tag{52}$$

is the pull-back of the dislocation density tensor to the current configuration, and

$$\tau^\kappa = \left(F^e m_0^\kappa\right) \cdot T\left(F^{e-T} n_0^\kappa\right) \tag{53}$$

is the resolved shear stress on the system $\kappa$. Constitutive equations for $V$ and $\dot{\gamma}^\kappa$ in terms of their respective driving forces may be specified by following the ideas in Section 3 with obvious adjustments for the changes in the driving forces.

In the second alternative for prescribing $\Pi$, we try to make a correspondence between dislocation dyads and the GND slipping distortion. At each point of the current configuration and at each instant of time we choose nine linearly independent dyads, say $E_r = b_r \otimes l_r$ (no sum), $r = 1,9$, not necessarily orthonormal (as dyads). These serve as a representation of dislocation types most likely to be mobile. For instance, for some crystal classes, such a choice can be made by ranking the most stressed slip systems and for each introducing two dislocation dyads corresponding to edge and screw types, while making sure that the collective set is linearly independent. Clearly, such a set of nine dyads forms a basis for the space of second order tensors and, under the standard trace inner product for tensors, admits a dual basis $\{E^r : r = 1,9\}$ that can be generated in the usual way from a fourth order metric tensor defined from the set $\{E_r : r = 1,9\}$. The tensor $\alpha^*$ can then be written as

$$\alpha^* = \sum_r \varpi^r E_r \quad ; \quad \varpi^r = \alpha^* : E^r \tag{54}$$

and introducing dislocation velocities $V_r$ corresponding to each dyad $E_r$, the GND slipping distortion is defined as

$$\Pi := \sum_r \varpi^r F^{e-1} E_r \times V_r. \tag{55}$$

The dissipation in the model

$$\int_C \sum_r \varpi^r X(TE_r) \cdot V_r \, dv + \int_C \sum_\kappa \tau^\kappa \dot{\gamma}^\kappa \, dv, \tag{56}$$

yields the driving force for the dislocation velocities $V_r$. Constitutive equations may now be specified in terms of driving forces, including preferred geometric constraints on the dislocation velocity directions (e.g. velocities constrained to lie on slip planes – but not necessarily, for instance, when modeling cross slip).

Initial and boundary conditions may be applied following the logic in the geometrically linear case (Sections 2.2, 2.3). When the GND slipping distortion is specified according to the second alternative (55), inflow boundary conditions on the GND slipping may be imposed for each dyadic component of $\boldsymbol{\alpha}^*$ at a boundary point utilizing an obvious adjustment of (19). The jump condition at an interface (grain boundary) with normal $N$ corresponding to $(48)_7$ is given by

$$[\![\boldsymbol{\Pi} + \tilde{\boldsymbol{L}}^p]\!] \times \boldsymbol{N} = \boldsymbol{0}, \tag{57}$$

where $[\![\ ]\!]$ refers to the difference of its argument, evaluated on the two sides of the interface. Due to the involvement of the lattice orientation in the definition of $\tilde{\boldsymbol{L}}^p$ at least, (57) depends on all five (kinematic) grain boundary parameters as well as the GND tensor and kinetics of SSD evolution in the crystals sharing the interface. As such, and even without further knowledge of physics/phenomenology related to how grain boundaries constrain the surface flow (17), just the five kinematic grain boundary parameters may be expected to have a significant effect on slip transmission at grain boundaries due to the imposition of (57) in calculations (e.g. when $(48)_7$ is imposed weakly).

The kinematical relationships between the fields $\tilde{\boldsymbol{\chi}}$ and $\boldsymbol{f}$ of the nonlinear model and $\boldsymbol{\chi}$ and $\boldsymbol{z}$ of the geometrically linear model are exposed by linearizing the formula for $\boldsymbol{F}^{e-1}$. Thus, let $\boldsymbol{x}_0$ represent position on the reference configuration (chosen to be the as-received body) and $\boldsymbol{x}$ a generic position on the current configuration. Utilizing

$$\boldsymbol{x} = \boldsymbol{x}_0 + \boldsymbol{u} \quad \text{and} \quad \frac{\partial(\ )}{\partial \boldsymbol{x}} = \frac{\partial(\ )}{\partial \boldsymbol{x}_0}\boldsymbol{F}^{-1} \approx \frac{\partial(\ )}{\partial \boldsymbol{x}_0}\left(\boldsymbol{I} - \frac{\partial \boldsymbol{u}}{\partial \boldsymbol{x}_0}\right) \approx \frac{\partial(\ )}{\partial \boldsymbol{x}_0}, \tag{58}$$

$\boldsymbol{F}$ being the deformation gradient and $\boldsymbol{I}$ the second-order identity, $(48)_4$ implies

$$\boldsymbol{F}^{e-1} = \tilde{\boldsymbol{\chi}} + \frac{\partial(\boldsymbol{f}-\boldsymbol{x})}{\partial \boldsymbol{x}} + \boldsymbol{I} \approx \tilde{\boldsymbol{\chi}} + \frac{\partial(\boldsymbol{f}-\boldsymbol{x}_0)}{\partial \boldsymbol{x}_0} - \frac{\partial \boldsymbol{u}}{\partial \boldsymbol{x}_0} + \boldsymbol{I}. \tag{59}$$

On the other hand, the assumption $\boldsymbol{F}^e - \boldsymbol{I} = \boldsymbol{U}^e$ is 'small' leads to the linearization

$$\boldsymbol{F}^{e-1} \approx \boldsymbol{I} - \boldsymbol{U}^e. \tag{60}$$

Setting (60) equal to (59) along with the definition $(23)_7$ we find

$$\begin{aligned}\tilde{\boldsymbol{\chi}} &= -\boldsymbol{\chi} \\ \frac{\partial(\boldsymbol{f}-\boldsymbol{x}_0)}{\partial \boldsymbol{x}_0} &= \frac{\partial \boldsymbol{z}}{\partial \boldsymbol{x}_0}.\end{aligned} \tag{61}$$

Up to a physically unimportant rigid translation, $\boldsymbol{f}$ is to be considered as the *plastic position vector* field determining the compatible part of the 'intermediate configuration' and $\boldsymbol{z}$ its *plastic displacement* field from the reference configuration. $\boldsymbol{\chi}$ may be considered as the incompatible part of $\boldsymbol{F}^e$ (with domain as the reference configuration) whereas $\tilde{\boldsymbol{\chi}}$ is the incompatible part of $\boldsymbol{F}^{e-1}$ by definition.

While the concept of a plastic distortion tensor, $\boldsymbol{F}^p$, with an associated multiplicative decomposition of the deformation gradient is unnecessary for our purposes, we nevertheless write it down to make contact with classical finite deformation plasticity (Lee, 1969):

$$\boldsymbol{F}^p := \boldsymbol{F}^{e-1}\boldsymbol{F} = \tilde{\boldsymbol{\chi}}\boldsymbol{F} + \frac{\partial \boldsymbol{f}}{\partial \boldsymbol{x}_0}. \tag{62}$$

## 6. Concluding Remarks

The primary goal of this work is an attempt to establish a mechanistically rigorous link between continuum plasticity theory and the theory of continuously distributed dislocations, with enough specificity so as to be able to be studied through numerical simulations. Our ultimate goal is to be able to study practical problems of mesoscopic and macroscopic plasticity from increasingly fundamental bases.

As shown in Part II of this paper, our model appears to perform reasonably. Some important theoretical issues that require attention is a detailed mathematical study of admissible boundary conditions for our model and, of course, a fundamental representation of the constitutive inputs for $\bar{\boldsymbol{V}}, \boldsymbol{L}^p$. While we are making progress on the latter issue, the former could greatly benefit from the attention of bona fide PDE theorists. Based on our numerical calculations, we are convinced that our nonlinear transport model is very rich from the mathematical point of view, and a precise understanding of issues like the effect of the physically motivated boundary conditions on the minimal space in which solutions should be expected (e.g. measure-valued solutions for $\boldsymbol{\alpha}$ that would naturally imply a probabilistic interpretation of results of our deterministic model), and the dependence of transitions in (in)stability of distinguished solutions (e.g. time-dependent spatially homogeneous solution) on geometric scale and overall strain would greatly benefit the practical application of the model to the prediction of evolution of microstructure and its effects on mechanical properties.

## 7. Acknowledgment

We thank Armand Beaudoin for his comments on the paper. Financial support for the work from the Program in Computational Mechanics of the U.S. Office of Naval Research, Grant No. N00014-02-1-0194 is gratefully acknowledged.